\documentclass[12pt, a4paper]{article}

\pdfoutput=1

\usepackage{amsmath}
\allowdisplaybreaks
\usepackage{amsfonts}
\usepackage{amssymb}
\usepackage{bbm}
\usepackage{verbatim}
\usepackage{dsfont}
\usepackage{booktabs}
\usepackage{slashed}

\usepackage{epsfig}
\usepackage{color}
\usepackage[table]{xcolor}
\usepackage{graphicx}
\usepackage[]{caption}
\usepackage[listofformat=empty,subrefformat=empty]{subfig}

\usepackage{float}
\usepackage{overpic}

\usepackage{enumerate}
\usepackage{hhline}
\usepackage{multirow}

\usepackage{cite}
\usepackage{xspace}
\usepackage{setspace}

\usepackage[pdftitle={Flux compactifications and naturalness},
  pdfauthor={},
  pdfsubject={},
  bookmarksopen, bookmarksnumbered, bookmarksopenlevel=2, colorlinks=false, linkcolor=blue, citecolor=blue, urlcolor=blue]{hyperref}

\renewcommand{\tfrac}{\genfrac{}{}{}1}
\newcommand{\ttfrac}{\genfrac{}{}{}2}

\newcommand{\pdb}{\partial_{\bar{z}}}
\newcommand{\pd}{\partial_{z}}
\newcommand{\vf}{\varphi}
\newcommand{\vfb}{\overline{\varphi}}

\begin{document}

\thispagestyle{empty}

\begin{flushright}
CPHT-RR020.042018\\
DESY 18-048\\
\end{flushright}
\vskip .8 cm
\begin{center}
{\Large {\bf Flux compactifications and naturalness}}\\[12pt]

\bigskip
\bigskip 
{
{\bf{Wilfried Buchmuller$^{a}$}\footnote{E-mail: wilfried.buchmueller@desy.de}},
{\bf{Markus Dierigl$^{b,c}$}\footnote{E-mail: m.j.dierigl@uu.nl}},  
{\bf{Emilian Dudas$^{d}$}\footnote{E-mail: emilian.dudas@cpht.polytechnique.fr}}
\bigskip}\\[0pt]
\vspace{0.23cm}
{\it $^{a}$ Deutsches Elektronen-Synchrotron DESY, 22607 Hamburg, Germany \\ \vspace{0.2cm}
$^b$ Institute for Theoretical Physics, Utrecht University, 3584 CC Utrecht,\\ The Netherlands \\ \vspace{0.2cm}
$^c$ Institute of Physics, University of Amsterdam, 1098 XH Amsterdam,\\ The Netherlands \\ \vspace{0.2cm}
$^d$ Centre de Physique Th\'eorique, \'Ecole Polytechnique, CNRS, Universit\'e Paris-Saclay, F-91128 Palaiseau, France}\\[20pt] 
\bigskip
\end{center}

\begin{abstract}
\noindent
Free massless scalars have a shift symmetry. This is usually broken by
gauge and Yukawa interactions, such that quantum corrections induce a
quadratically divergent mass term. In the Standard Model this leads to
the hierarchy problem of the electroweak theory, the question why the
Higgs mass is so much smaller than the Planck mass. We present an example where a large scalar mass term is avoided by coupling the scalar to an
infinite tower of massive states which are obtained from a
six-dimensional theory compactified on a torus with magnetic flux. 
The series of divergent quantum corrections adds up to zero, and we
show explicitly that the shift symmetry of the scalar is preserved in
the effective four-dimensional theory despite the presence of gauge and Yukawa interaction terms.
\end{abstract}

\newpage 
\setcounter{page}{2}
\setcounter{footnote}{0}

{\renewcommand{\baselinestretch}{1.5}

\section{Introduction}
\label{sec:Introduction}

Compactifications on tori with magnetic flux play an important role in
string theories and higher-dimensional field theories (see, for example,
  \cite{Angelantonj:2002ct,Blumenhagen:2006ci,Ibanez:2012zz}). Due to
the index theorem they lead to a multiplicity of chiral fermions,
which can be used to explain the number of quark-lepton generations
\cite{Witten:1984dg}. Moreover, magnetic flux is an important source
of supersymmetry breaking \cite{Bachas:1995ik}. Magnetic
compactifications of higher-dimensional field theories have been 
thoroughly studied in Ref.~\cite{Cremades:2004wa}. These results have
been used to construct
interesting supersymmetric models of particle physics and to compute
Yukawa couplings (see, for example,
\cite{Cremades:2004wa,Kobayashi:2010an,Hamada:2012wj}).
Making use
of flux configurations that break supersymmetry one can also construct
extensions of the Standard Model with high-scale supersymmetry 
\cite{Buchmuller:2015jna}.

The components of higher-dimensional gauge fields along compact
dimensions play a special role for compact spaces with non-trivial
topology. Their zero modes, often called Wilson-line (WL) scalars
are interesting
candidates for Higgs fields in four dimensions (4d)
\cite{Hosotani:1983xw,Hatanaka:1998yp,Dvali:2001qr}. Compactifying a
five-dimensional (5d) theory on a circle, or a six-dimensional (6d)
theory on a torus, one finds a discrete set of large gauge transformations
in the 4d theory,
due to the higher-dimensional gauge invariance and the non-trivial 
topology of the compact manifold.
These large gauge transformations act as discrete
shifts on WL scalars and can therefore protect their masses
from quadratic divergencies. Identifying Higgs fields as WL scalars,
one obtains finite Higgs masses, determined by size of the extra
dimensions, $m_H^2 \propto L^{-2}$, where $L$
is a typical length scale of the internal space \cite{ArkaniHamed:2001nc,
Antoniadis:2001cv,Alfaro:2006is}. This mechanism to protect Higgs
masses is of interest in scenarios with large extra dimensions, where
the scale of electroweak symmetry breaking is tied to the size of the
compact space.

How can one protect scalar masses if the ultraviolet cutoff of the 
theory lies much above the scale of electroweak symmetry breaking? 
In this case the only known candidate
for a protection mechanism is a continuous shift symmetry, like the Peccei-Quinn
symmetry \cite{Peccei:1977hh} in axion physics or the shift symmetry of a Goldstone
boson in composite Higgs models (see, for example,
\cite{Panico:2015jxa}). Such a
mechanism would be needed in models of high-scale supersymmetry, like
the one considered in \cite{Buchmuller:2015jna}. 
In the following we shall give an example that illustrates how
such a shift symmetry can indeed arise in 
compactifications with magnetic flux, and we shall identify the
higher-dimensional origin of the symmetry.

In Ref.~\cite{Buchmuller:2016gib} we have worked out the magnetic
compactification of a supersymmetric 6d Abelian gauge theory on a torus, and we
have compared the results with the standard compactification without
flux. In the latter case bosonic and fermionic one-loop corrections to
the mass of the WL scalar are separately finite
and cancel each other due to unbroken supersymmetry. On the contrary,
in the case of flux compactification, bosonic and fermionic
contributions are zero separately, once the complete tower of massive
states is taken into account. We argued that this surprizing cancellation  
of an infinite number of terms is a consequence of a 6d symmetry, the invariance
under translations on the torus under which the WL scalar
transforms with a shift. The vanishing of the one-loop corrections to the mass of the WL scalar has subsequently been carefully studied in \cite{Ghilencea:2017jmh}.

Notice that our field theory setup was widely studied in string theory compactifications with internal magnetic fields \cite{Bachas:1995ik,Angelantonj:2000hi} and in the T-dual version of D-branes at angles (or intersecting brane models) \cite{Berkooz:1996km, Blumenhagen:2000wh, Aldazabal:2000cn}, as a way to partially or completely break supersymmetry and to induce fermion chirality. However, a field theory approach has its own advantages, namely,
more flexibility in searching for realistic models of particle physics and the avoidance of technical difficulties with quantum corrections for string theory models with broken supersymmetry, 
see e.g.~\cite{Dudas:2004nd}. 

In this paper we study the cancellation of loop corrections to the
mass of the WL scalar in more detail. Since the cancellation
is independent of supersymmetry, we focus on the simplest possible
model, a single 6d Weyl fermion interacting with an Abelian gauge field.
In Section~2 we provide details of the flux compactification on a
torus with emphasis on the symmetries of the 6d theory and the
couplings of the tower of massive states in the effective 4d theory. 
Quantum corrections to the mass of the WL scalar are discussed in
Section~3. We first recall the cancellations at one-loop order once
the tower of massive states is taken into account. We then show that
the 4d action possesses an exact shift symmetry, including the 
couplings to all massive states. In Section~4, we summarize
our results and discuss the prospects to extend the presented model
to chiral Higgs models. The connection between the considered field theory and quantum mechanics on a magnetized torus is discussed in the appendix.

\section{Flux compactification on a torus}
\label{sec:torusflux}

Let us now consider a  left-handed 6d Weyl fermion interacting with an Abelian
gauge field,\footnote{In the following we ignore anomalies. Note that our discussion will not change for an anomaly-free fermion spectrum.}
\begin{align} \label{S6}
S_6 = \int d^6 x \left( -\frac{1}{4} F^{MN} F_{MN} +
i\overline{\Psi}\Gamma^M D_M \Psi \right)\,,
\end{align}
where $D_M = \partial_M + iq A_M$, $M=0,\ldots 6$, $F_{MN}
= \partial_M A_N - \partial_N A_M$ and $\Gamma_7 \Psi = - \Psi$.
The 6d space is a product of 4d Minkowski space and a square torus
$T^2$ of area $L^2$. It is convenient to decompose the 6d Weyl spinor
into two independent two-component Weyl spinors $\psi$ and $\chi$. For gamma matrices
in the Weyl basis, one has\footnote{We follow the conventions in
 Ref.~\cite{Wess:1992cp}. Our 6d gamma matrices satisfy the algebra
  $\{\Gamma_M,\Gamma_N\} = -2\eta_{MN}$, with $\text{diag}(\eta_{MN})
  = (-1,+1,\ldots,+1)$.}
\begin{align}
\Psi = \begin{pmatrix} \psi_L \\ \psi_R \end{pmatrix}:\quad
\gamma_5 \psi_L = - \psi_L\,, \quad \gamma_5 \psi_R = \psi_R\,, \\
\psi_L = \begin{pmatrix} \psi \\ 0 \end{pmatrix} \,, \quad
\psi_R = \begin{pmatrix} 0 \\ \overline{\chi} \end{pmatrix} \,.
\end{align}
The Weyl fermions $\psi$ and $\chi$ have charges $q$ and $-q$,
respectively, and the fermionic part of the action \eqref{S6} reads
\begin{align}\label{S6df} 
S_{6f} 
&= \int d^6 x \Big(-i\psi\sigma^\mu \overline{D}_\mu \overline{\psi} - i\chi\sigma^\mu D_\mu \overline{\chi}\nonumber\\
&\hspace*{2cm} -\chi\left(\pd + \sqrt{2} q \phi\right) \psi 
- \overline{\chi}\left(\pdb + \sqrt{2} q \overline{\phi}\right) \overline{\psi} \Big)\,,
\end{align}
where $D_\mu = \partial_\mu + iqA_\mu$, $\overline{D}_\mu
= \partial_\mu - iqA_\mu$ and
\begin{align}
\phi = \frac{1}{\sqrt{2}}\left(A_6 + i A_5\right)\,, \quad
z = \frac{1}{2}\left(x_5 + i x_6\right)\,, \quad \pd = \partial_5 -
i \partial_6\,.
\end{align}
The coordinates take values in the interval $x_{5,6} \in [0,L)$. In
the following we set $L=1$.
The gauge kinetic term can be expressed in terms of the fields $A_\mu$ and $\phi$,
\begin{align}\label{S6dg}
S_{6g} &= \int d^6 x \left(-\frac{1}{4} F^{MN}F_{MN}\right) \nonumber\\ 
&= \int d^6 x \Big( -\frac{1}{4}F^{\mu\nu} F_{\mu\nu} 
- \partial^\mu \overline{\phi}\partial_\mu \phi
- \frac{1}{4}\left(\pd \overline{\phi} + \pdb \phi\right)^2 \nonumber\\
&\hspace*{2cm} - \frac{1}{2}\pdb A^\mu \pd A_\mu - \frac{i}{\sqrt{2}}\partial_\mu A^\mu
\left(\pd \overline{\phi} - \pdb\phi\right) \Big)\,.
\end{align} 
Constant magnetic flux in the compact dimensions corresponds to a
vacuum configuration. For $\langle A_5\rangle = - \tfrac{1}{2} f x_6$, $\langle
A_6\rangle = \tfrac{1}{2} fx_5$, corresponding to 
$\langle \phi\rangle =  \frac{1}{\sqrt{2}} f\bar{z}$, the vacuum field equations are satisfied,\footnote{Note
  that this non-trivial gauge background requires the introduction of four patches on the torus.}
\begin{align}
\pd \left(\pd \langle\overline{\phi}\rangle + \pdb \langle\phi\rangle\right) = 0\,.
\end{align}
The magnetic flux is quantized in units of the torus area,
\begin{align}
\frac{q}{2\pi} \int_{T^2} F = \frac{q}{2\pi} f = N \in \mathbb{Z} \,.
\end{align}
Shifting the scalar field $\phi$ around the flux background,
\begin{align} 
\phi =  \frac{f}{\sqrt{2}} \bar{z} + \vf\,,
\end{align}
the 6d action takes the form of Eqs.~\eqref{S6df} and \eqref{S6dg}, with
$\phi$ replaced by $\vf$, up to a cosmological
constant\footnote{Once gravity is included, the backreaction of the
  cosmological term on the compact manifold has to be taken into
  account. However, we have applications in mind with $f \ll M_{\rm
    PL}$, where gravitational corrections are expected to be small.} and a flux-dependent bilinear term of the
Weyl fermions,
\begin{align}\label{S6shifted}
S_{6} &= \int d^6 x \Big( -\frac{1}{4}F^{\mu\nu} F_{\mu\nu} 
- \partial^\mu \overline{\vf}\partial_\mu \vf
- \frac{1}{4}\left(\pd \overline{\vf} + \pdb \vf\right)^2  -
\frac{1}{2} f^2\nonumber\\
&\hspace*{2cm} - \frac{1}{2}\pdb A^\mu \pd A_\mu - \frac{i}{\sqrt{2}}\partial_\mu A^\mu
\left(\pd \overline{\vf} - \pdb\vf\right) \nonumber\\
&\hspace*{2cm}-i\psi\sigma^\mu \overline{D}_\mu \overline{\psi} - i\chi\sigma^\mu D_\mu \overline{\chi}\nonumber\\
&\hspace*{2cm} -\chi\left(\pd + qf\bar{z} + \sqrt{2} q \vf\right) \psi 
-\overline{\chi}\left(\pdb + qf z + \sqrt{2} q \vfb\right) \overline{\psi} \Big)\,.
\end{align}
The action is invariant under translations on the torus, which act in
the standard way as $\delta_T =
\epsilon\pd + \overline{\epsilon}\pdb$ on the fields $A_\mu$, $\psi$
and $\chi$. The breaking of translational invariance by the background
gauge field can be compensated by a shift of $\vf$,
\begin{align}\label{deltaT}
\delta_T \vf = \left(\epsilon\pd + \overline{\epsilon} \pdb\right)\vf 
+  \frac{\bar{\epsilon}}{\sqrt{2}} f\,.
\end{align}
The Lagrangian density in \eqref{S6shifted} then transforms into a
total divergence.\footnote{To prove the invariance of the action one
  has to take into account that gauge field and charged fermions are
  fiber bundles defined on four patches of the torus.}
Furthermore, the action is invariant with respect to the
following local 6d transformation,
\begin{align}
 \vf_\Lambda = \vf  - \frac{1}{\sqrt{2}}\pd\Lambda\,,\quad
\psi_\Lambda = e^{q\Lambda}\psi\,,\quad \chi_\Lambda = e^{- q\Lambda}\chi\,,\quad
\Lambda = f\left(\alpha\bar{z} - \bar{\alpha}z\right)\,, \label{shifttrafo}
\end{align}
where $\alpha$ is a complex parameter. Such transformations have first
been considered in \cite{Scherk:1978ta}. Note that they change the boundary
conditions of the fermion wave functions.
For infinitesimal $\alpha$ the transformation reads
\begin{align}\label{deltaL}
\delta_\Lambda \vf = -\frac{1}{\sqrt{2}}\pd\Lambda\,,\quad
\delta_\Lambda \psi = q\Lambda\psi\,,\quad \delta_\Lambda \chi = - q\Lambda\chi\,.
\end{align}
For the complex scalar field, the
gauge transformation corresponds to a shift,
\begin{align}\label{deltaLphi}
\delta_\Lambda \vf = \frac{\bar{\alpha}}{\sqrt{2}} f \,.
\end{align} 
In order to obtain the effective 4d action one expands the 6d fields
into mode functions corresponding to eigenstates of the kinetic term
of the compact dimensions. For charged fields these are Landau levels
obtained from an harmonic oscillator algebra \cite{Bachas:1995ik,Alfaro:2006is,Braun:2006se}. The
identification of annihilation and creation operators depends on the
sign of $qf$. Without loss of generality we choose $qf > 0$. There are two pairs of
annihilation and creation operators
\begin{align}
\quad &a_+ =\frac{i}{\sqrt{2qf}} \left(\pd + qf\bar{z}\right)\,, \quad
a_+^{\dagger} =\frac{i}{\sqrt{2qf}} \left(\pdb - qfz\right)\,,\label{aplus}\\
\quad &a_- =\frac{i}{\sqrt{2qf}} \left(\pdb + qfz\right)\,, \quad
a_-^{\dagger} =\frac{i}{\sqrt{2qf}} \left(\pd - qf\bar{z}\right)\,. \label{aminus}
\end{align}
They satisfy the commutation relations
$[a_\pm,a^\dagger_\pm]=1$, $[a_\pm,a_\mp]=0$,
$[a_\pm,a^\dagger_\mp]=0$. In terms of the annihilation and creation
operators the mass-square operators for the fermions $\psi$ with
charge $+q$ and $\chi$ with charge $-q$ are given by
\begin{align}\label{massoperator}
\mathcal{M}^2_+ = 2qf a_+^\dagger a_+\,,\quad 
 \mathcal{M}^2_- = 2qf \left(a_-^\dagger a_- + 1\right)\,. 
\end{align}
The ground state wave functions are determined by
\begin{align}
a_+ \xi_{0,j} = 0\,, \quad a_- \overline{\xi}_{0,j}=0\,,
\end{align}
where $j = 0,\ldots |N|-1$ labels the degeneracy of the ground
state. An orthonormal set of higher mode functions
is given by
\begin{align} 
\xi_{n,j} = \frac{i^n}{\sqrt{n!}} \left(a_+^\dagger\right)^n \xi_{0,j}\,,\quad
\overline{\xi}_{n,j} = \frac{i^n}{\sqrt{n!}}
\left(a_-^\dagger\right)^n \overline{\xi}_{0,j}\,.
\end{align}
Annihilation and creation operators act on these mode functions as
\begin{align}
a_+ \xi_{n,j} &= i \sqrt{n}\ \xi_{n-1,j}\,, \quad a_+^\dagger \xi_{n,j}
= -i\sqrt{n+1}\ \xi_{n+1,j}\,,\label{axiplus}\\
a_- \overline{\xi}_{n,j} &= i \sqrt{n}\ \overline{\xi}_{n-1,j}\,, \quad a_-^\dagger \overline{\xi}_{n,j}
= -i\sqrt{n+1}\ \overline{\xi}_{n+1,j}\,,\label{aximinus}
\end{align}
and the mode expansions of the fermion fields $\psi$ and $\chi$ with
charges $+q$ and $-q$, respectively, read 
\begin{align}\label{matterKK}
\psi = \sum_{n,j} \psi_{n,j}\xi_{n,j}\,,\quad \chi = \sum_{n,j}
\chi_{n,j}\overline{\xi}_{n,j} \,.
\end{align}
Since the gauge fields $A_\mu$ and $\vf$ do not feel the flux, they
have an expansion in terms of standard Kaluza-Klein modes. The theory
has a number of 4d zero modes. According to Eq.~\eqref{massoperator}, 
and in accord with the index theorem, there are
$|N|$ left-handed fermionic zero modes $\psi_{0,j}$. Moreover, there will be
zero modes $A_{0\mu}$ due to 4d gauge invariance, and, up to
quantum corrections, a massless complex scalar $\vf_0$\footnote{Note that in a supergravity extension of the present model the zero modes $A_{0\mu}$ become massive due to the
Stueckelberg mechanism. For a recent discussion of the
  interplay of flux and the Green-Schwarz mechanism, see
  \cite{Buchmuller:2015eya}.}. 
The action for the zero mode $\vf_0$, $A_{0\mu}$,
and the matter fields is easily obtained by inserting the expansions
\eqref{matterKK} into the action \eqref{S6shifted}. The result reads
\begin{align}\label{S4phi0}
S_{4} &= \int d^4 x \Big( - \partial^\mu \overline{\vf}_0\partial_\mu \vf_0
+ \sum_{n,j}\Big(
-i\psi_{n,j}\sigma^\mu \overline{D}_\mu \overline{\psi}_{n,j} - i\chi_{n,j}\sigma^\mu D_\mu \overline{\chi}_{n,j}\nonumber\\
&\hspace*{2cm} -\sqrt{2qf(n+1)}\chi_{n,j}\psi_{n+1,j} -
\sqrt{2}q\vf_0 \chi_{n,j}\psi_{n,j} + \text{h.c.} \Big)\Big)\,.
\end{align}
This is the fermionic part of the supersymmetric action derived in
\cite{Buchmuller:2016gib}. It describes $|N|$ left-handed fermions $\psi_{L j}$ and an infinite tower of massive Dirac fermions
$\Psi_{n,j}$,
\begin{align}
\psi_{L j} = \begin{pmatrix} \psi_{0,j} \\ 0 \end{pmatrix} \,, \quad
\Psi_{n,j} = \begin{pmatrix} \psi_{n+1,j} \\
  \overline{\chi}_{n,j} \end{pmatrix} \,, 
\end{align}
which interact via Yukawa couplings with a massless scalar,
\begin{align}\label{S4dirac}
S_{4} &= \int d^4 x \Big( - \partial^\mu \overline{\vf}_0 \partial_\mu \vf_0
 + \sum_{n,j}\Big( i\overline{\psi}_{L j}\gamma^\mu D_\mu \psi_{L j} 
+ i\overline{\Psi}_{n,j}\gamma^\mu D_\mu \Psi_{n,j} \nonumber\\
&\hspace*{1cm} + \sqrt{2qf(n+1)}\ \overline{\Psi}_{n,j}\Psi_{n,j} 
 + \sqrt{2}q\vf_0\left(\overline{\Psi}_{0,j}\tfrac{1-\gamma_5}{2}\psi_{L j} 
+ \overline{\Psi}_{n+1,j}\tfrac{1-\gamma_5}{2}\Psi_{n, j}\right) \nonumber\\
&\hspace*{1cm}+ \sqrt{2}q\vfb_0\left(\overline{\psi}_{Lj}\tfrac{1+\gamma_5}{2}\Psi_{0, j} 
+ \overline{\Psi}_{n,j}\tfrac{1+\gamma_5}{2}\Psi_{n+1, j}\right) 
\Big)\Big)\,.
\end{align}
Note that the scalar $\vf_0$ couples to different mass
eigenstates. Integrating out the heavy fermions $\Psi_{n,j}$ yields
the effective low energy action of the zero modes $\vf_0$ and
$\psi_{Lj}$. The first contribution is due to the exchange of
$\Psi_{0,j}$. Expanding the propagator as
\begin{align}
\langle\Psi_{0,j}(x)\overline{\Psi}_{0,j'}(x')\rangle =
-\frac{i}{m_0}\left(1+\frac{i\gamma^\mu\partial_\mu}{m_0} + \ldots
\right)\delta_{j,j'}\delta^4(x-x')\,,
\end{align}
with $m_0 = \sqrt{2qf}$, one obtains 
\begin{align}\label{S4effective}
S_{eff} = \int d^4 x \Big( &- \partial^\mu \overline{\vf}_0 \partial_\mu \vf_0
 + i\overline{\psi}_{L j}\gamma^\mu D_\mu \psi_{L j}\nonumber\\
&+i\frac{q^2}{m_0^2}\overline{\psi}_{L,j}\gamma^\mu\psi_{L,j}
\left(\vfb_0\partial_\mu\vf_0 - \partial_\mu\vfb_0 \vf_0 \right) + \ldots\Big)\,,
\end{align}
where we have used the equation of motion for $\psi_{Lj}$ to leading
order, i.e. $\gamma^\mu\partial_\mu\psi_{Lj} = 0$. One easily
verifies that this effective action is invariant under a constant
shift of $\vf_0$.

The effective action \eqref{S4effective} is very different from the 4d
action without magnetic flux. In this case one obtains a vector-like
theory, and after spontaneous symmetry breaking the lowest states
of the spectrum consist of a Dirac fermion, a real scalar and a
vector, which all have masses of the order of the compactification scale. No massless
states are left. On the contrary, the action \eqref{S4effective} does
contain massless chiral fermions and a WL scalar which is kept massless by
 a continuous shift symmetry. However, contrary to the case of the Standard Model, its vacuum expectation value does not give mass to the chiral fermions.

\vspace{-0.4cm}
\section{Quantum corrections and shift symmetry}

In general, Yukawa interactions violate the shift symmetry of a free
massless scalar, and as a consequence quantum corrections generate a
mass term. Indeed, keeping the lightest massive fermion $\Psi_{0,j}$ in
addition to the zero modes $\Psi_{Lj}$, one obtains from the standard
one-loop diagrams (see Figure~1, left),
\begin{align}
\delta m^2_{\vf_0} &= -2 q^2 |N| \int \frac{d^4k}{(2\pi)^4} \frac{2 k^2}{k^2\left(k^2 + 2qf\right)} \nonumber\\
&= - \frac{q^2 |N|}{4\pi^2}\left(\Lambda^2 - 2qf
  \ln \Big(\frac{\Lambda^2}{2qf}\Big) + \ldots \right)\,,
\end{align}
where we have introduced a momentum cutoff $\Lambda$ as
regulator. Usually, the quadratic divergence is removed by a
counter term, leaving an undetermined finite mass for the scalar $\vf_0$. 
\begin{figure}
\begin{center} 
\includegraphics[width = 0.4\textwidth]{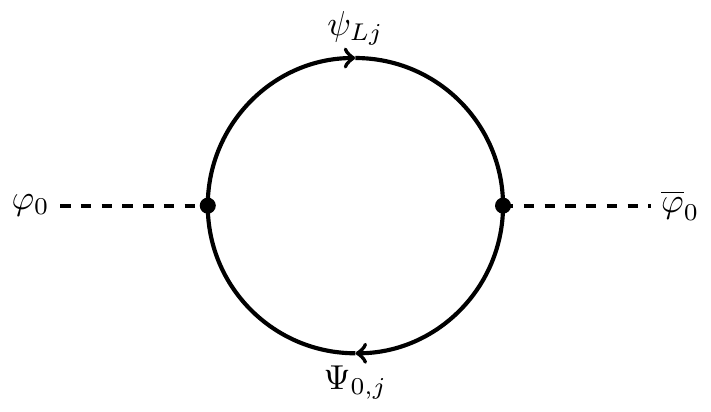}
\hspace{0.7cm}\includegraphics[width = 0.4\textwidth]{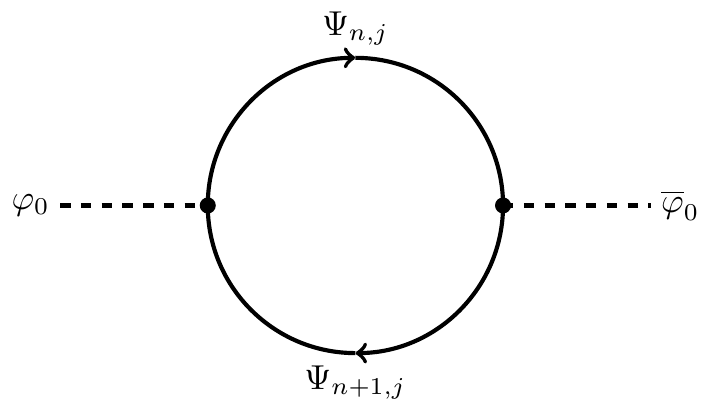}
\end{center}
\vspace{0.3cm}	
\caption{One-loop contributions to the scalar mass term. Left:
  contribution of $\psi_{Lj}$ and $\Psi_{0,j}$; right: contributions
of the massive fermions $\Psi_{n,j}$ and $\Psi_{n+1,j}$.}
	\label{phi2fer}
\end{figure}
In Ref.~\cite{Buchmuller:2016gib} it was shown that the situation drastically changes once
the Yukawa couplings to the entire tower of massive states are taken
into account (see Figure~1, right). One then obtains  
\begin{align}
\delta m_{\vf_0}^2 &= -2 q^2 |N| \sum_n \int \frac{d^4k}{(2\pi)^4} \frac{2
  k^2}{\left(k^2 + 2qf n \right)\left(k^2 + 2qf (n + 1)\right)}
\nonumber\\
&= 4 q^2 |N| \sum_n \int \frac{d^4 k}{(2 \pi)^4} \left(
  \frac{n}{k^2 + 2qf n} - \frac{n+1}{k^2 + 2qf (n + 1)}\right)\,.
\end{align}
Using the Schwinger representation of the propagators, performing the
momentum integrations and interchanging $t$-integration and summation,
one finds 
\begin{align}
\delta m^2_{\vf_0}  &= \frac{q^2}{4 \pi^2} |N| \sum_n \int_0^{\infty} dt \, \frac{1}{t^2} \left( n e^{- 2qf n t} - (n+1) e^{- 2qf (n + 1) t}\right) \nonumber\\
&=  \frac{q^2}{4 \pi^2} |N| \int_0^{\infty} dt \, \frac{1}{t^2} \left( \frac{e^{2qf t}}{(e^{2qf t} - 1)^2} - \frac{e^{2qf t}}{(e^{2qf t} - 1)^2} \right) \nonumber\\
&= \ \, 0 \,.
\end{align}
To obtain this remarkable cancellation it is crucial to perform the
summation before the momentum integration, as in
Ref.~\cite{Antoniadis:2001cv}. In this way the symmetries
of the gauge theory in the compact dimensions are preserved. 

What is the origin of the cancellation of the quantum corrections to
the scalar mass term and can one understand it at the level of the
four-dimensional theory? As discussed in the previous section the
six-dimensional theory is invariant under translations, which include
a shift of the scalar field $\vf_0$. The generators of the
translations, $\partial_z$ and $\partial_{\bar{z}}$, do not commute with
the mass-squared operators $\mathcal{M}^2_\pm$. However, the mode functions
are eigenfunctions of $\mathcal{M}^2_\pm$. Therefore, they have no simple transformation
law under the action of $\partial_z$ and $\partial_{\bar{z}}$. Instead, the
whole tower is reshuffled. A simple transformation of the mode function can be
obtained by combining translations with the transformation
$\delta_\Lambda$, Eq.~\eqref{deltaL}, as follows:
\begin{align}
\delta\psi &= \left(\delta_T + \delta_{\Lambda,\alpha=\epsilon}\right)\psi \nonumber\\
&= \left(\epsilon\pd + \bar{\epsilon}\pdb + qf(\epsilon\bar{z} -
  \bar{\epsilon}z)\right)\psi \nonumber \\
&= -i\sqrt{2qf}(\epsilon a_+ + \bar{\epsilon} a_+^\dagger) \psi\,. \label{trafopsi}
\end{align}
Clearly, this infinitesimal transformation only connects mode functions of
neighboring mass eigenvalues. As we show in Appendix~\ref{app:qm}, this symmetry also manifests itself in the quantum mechanical analysis of a charged particle on a magnetized torus.
Using Eqs.~\eqref{axiplus} one obtains
\begin{align}
\delta\psi &= -i\sqrt{2qf}
\sum_{n,j} \psi_{n,j}(\epsilon a_+ + \bar{\epsilon} a_+^\dagger)\xi_{n,j} 
= \sum_{n,j} \delta\psi_{n,j}\xi_{n,j}\,,\nonumber\\
\delta\psi_{n,j} &= \sqrt{2qf}(\epsilon\sqrt{n+1}\ \psi_{n+1,j} -
  \bar{\epsilon}\sqrt{n}\ \psi_{n-1,j})\,.\label{deltapsi}
\end{align}
Analogously, the transformation of the matter field $\chi$ with charge
$-q$ is given by
\begin{align}
\delta\chi &= \left(\delta_T + \delta_{\Lambda,\alpha=\epsilon}\right) \chi \nonumber\\
&=(\epsilon\pd + \bar{\epsilon}\pdb - qf(\epsilon\bar{z} -
  \bar{\epsilon}z))\chi \nonumber \\
&= -i\sqrt{2qf}(\epsilon a_-^\dagger + \bar{\epsilon}
  a_-) \chi\,. \label{trafochi}
\end{align}
Using Eqs.~\eqref{aximinus} one finds
\begin{align}
\delta\chi &= -i\sqrt{2qf}
\sum_{n,j} \chi_{n,j}(\epsilon a_-^\dagger + \bar{\epsilon}
  a_-)\overline{\xi}_{n,j} 
= \sum_{n,j} \delta\chi_{n,j}\overline{\xi}_{n,j}\,,\nonumber\\
\delta\chi_{n,j} &= \sqrt{2qf}(-\epsilon\sqrt{n}\ \chi_{n-1,j} +
  \bar{\epsilon}\sqrt{n+1}\ \chi_{n+1,j})\,. \label{deltachi}
\end{align}
Given the transformation laws \eqref{deltapsi} and \eqref{deltachi}
it is straightforward to verify the invariance of the action \eqref{S4phi0}. For
instance, for the Yukawa term one has
\begin{align}
\delta\Big(\sum_{n,j} \chi_{n,j}\psi_{n,j}\Big) &= \sqrt{2qf}\sum_{n,j} \Big(-\epsilon\sqrt{n}\ \chi_{n-1,j} \psi_{n,j}+
  \bar{\epsilon}\sqrt{n+1}\ \chi_{n+1,j}\psi_{n,j} \nonumber\\
&\hspace{2cm}+\epsilon\sqrt{n+1}\ \chi_{n,j}\psi_{n+1,j} -
  \bar{\epsilon}\sqrt{n}\ \chi_{n,j}\psi_{n-1,j}\Big)=\ 0\,.
\end{align}
\vspace{-0.2cm}
Similarly, also the fermion kinetic terms are invariant. The remaining
part is
\begin{align}
\delta S_{4} &\supset \int d^4 x \Big( - \partial^\mu\delta\overline{\vf}_0\partial_\mu \vf_0
- \partial^\mu \overline{\vf}_0\partial_\mu \delta\vf_0\\
&\hspace*{1.5cm} + \sum_{n,j}\big(-\sqrt{2qf(n+1)}\delta \left(\chi_{n,j}\psi_{n+1,j}\right) 
-\sqrt{2}q\delta\vf_0 \chi_{n,j}\psi_{n,j} + \text{h.c.} \big)\Big)\,.\nonumber
\end{align}
\vspace{-0.2cm}
The variation of the mass term reads
\begin{align}
\delta\Big(\sum_{n,j} \sqrt{n+1}\chi_{n,j}\psi_{n+1,j}\Big) 
&= \sqrt{2qf}\sum_{n,j} \sqrt{n+1}\Big(-\epsilon\sqrt{n}\
\chi_{n-1,j} \psi_{n+1,j}\nonumber\\
&\hspace{-2.5cm}+\bar{\epsilon}\sqrt{n+1}\ \chi_{n+1,j}\psi_{n+1,j} 
+\epsilon\sqrt{n+2}\ \chi_{n,j}\psi_{n+2,j}
-\bar{\epsilon}\sqrt{n+1}\ \chi_{n,j}\psi_{n,j}\Big)\nonumber\\
&= -\bar{\epsilon} \sqrt{2qf}\sum_{n,j} \chi_{n,j}\psi_{n,j}\,,
\end{align}
\vspace{-0.3cm}
which yields for the 4d action
\begin{align}
\delta S_{4} &\supset \int d^4 x \Big( - \partial^\mu\delta\overline{\vf}_0 \partial_\mu \vf_0
- \partial^\mu \overline{\vf}_0 \partial_\mu \delta\vf_0 \nonumber\\
&\hspace*{2cm} + (2qf \bar{\epsilon} - \sqrt{2}q\delta\vf_0)\ \sum_{n,j}\chi_{n,j}\psi_{n,j} + \text{h.c.} \Big)\,.
\end{align}
Hence, the action is invariant if the scalar $\vf_0$ transforms as
\begin{align}\label{deltaphi0}
\delta\vf_0 = \sqrt{2} \bar{\epsilon} f \,.
\end{align}
This is precisely the shift inferred from the two transformation laws
of the 6d field $\vf$ given in Eqs.~\eqref{deltaT} and \eqref{deltaLphi}.
\begin{figure}
\begin{center} 
\includegraphics[width = 0.4\textwidth]{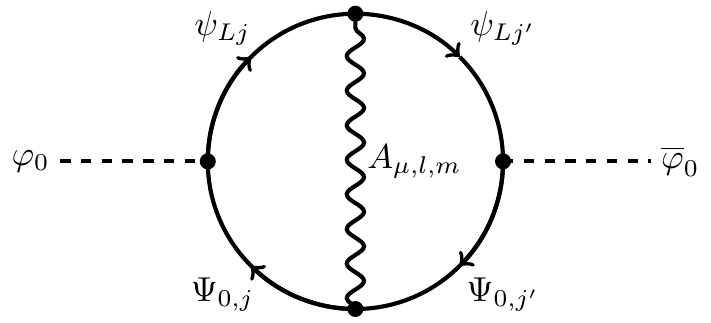}
\hspace{0.7cm}\includegraphics[width = 0.4\textwidth]{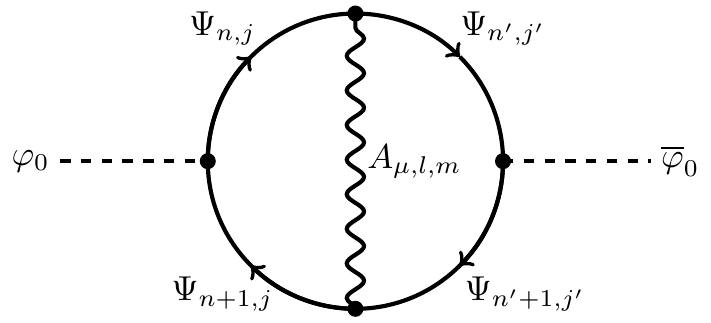}
\end{center}
\vspace{0.3cm}	
\caption{Two-loop contributions to the scalar mass term
which involve Kaluza-Klein modes of the vector field.}
	\label{phi2fer}
\end{figure}

So far we have discussed the coupling of the zero mode $\vf_0$ to the
matter fields $\psi_{n,j}$ and $\chi_{n,j}$, and we have seen that at
one-loop order no mass term is generated. However, the full theory
also contains the Kaluza-Klein excitations of the gauge fields $A_\mu$
and $\vf$, which enter at higher loop order (see Figure~2), and it is an important
question whether also these corrections preserve the shift symmetry of
$\vf_0$. Splitting the gauge fields into zero modes and KK
excitations, one has
\begin{align}
\vf &= \vf_0 + \vf'\,, \quad A_\mu = A_{0\mu} + A'_\mu\,,\nonumber\\
\vf' &= \sum_{l,m}\vf_{l,m}\lambda_{l,m}\,,\quad
A'_\mu = \sum_{l,m}A_{\mu,l,m}\lambda_{l,m}\,,\label{gaugeKK}
\end{align}
with zero modes ($l=m=0$) excluded in the sum, and with the
standard orthonormal mode functions
\begin{align}
\lambda_{l,m} &= e^{zM_{l,m} - \bar{z} \overline{M}_{l,m}} = \overline{\lambda}_{-l,-m} \,,\quad M_{l,m} =
2\pi (m+il)\,.
\end{align}
Since the vector field is real, one has $A_{\mu,l,m} =
A_{\mu,-l,-m}$. For the gauge fields the mass-squared operator is
given by $\mathcal{M}^2 = -\partial_{\bar{z}}\partial_z$, which
commutes with the generators of translations. In fact, they are
eigenfunctions of $\partial_{\bar{z}}$ and $\partial_z$. Hence, a simple
transformation law is obtained for the transformation $\delta$,
\begin{align}
\delta\vf' &= (\epsilon \partial_z + \bar{\epsilon}\partial_{\bar{z}})\vf'
= \sum_{l,m}\delta\vf_{l,m}\lambda_{l,m}\,,\\
\delta A'_\mu &= (\epsilon \partial_z + \bar{\epsilon}\partial_{\bar{z}})A'_\mu
= \sum_{l,m}\delta A_{\mu,l,m}\lambda_{l,m}\,,
\end{align}
which yields the transformation law of the mode functions 
\begin{align}
\delta\vf_{l,m} &= (\epsilon M_{l,m} - \bar{\epsilon}\overline{M}_{l,m})\vf_{l,m}\,,\label{deltaphiprime}\\
\delta A_{\mu,l,m} &= (\epsilon M_{l,m} - \bar{\epsilon}\overline{M}_{l,m})A_{\mu,l,m}\,.\label{deltaAprime}
\end{align}
These equations together with Eqs.~\eqref{deltapsi}, \eqref{deltachi} and \eqref{deltaphi0},
\begin{align}
\delta \vf_0 &= \sqrt{2} \overline{\epsilon} f \,, \nonumber\\
\delta \psi_{n,j} &= \sqrt{2qf} (\epsilon \sqrt{n+1} \, \psi_{n+1,j} - \overline{\epsilon} \sqrt{n} \, \psi_{n-1,j}) \,, \nonumber\\
\delta \chi_{n,j} &= \sqrt{2qf} (- \epsilon \sqrt{n} \, \chi_{n-1,j} + \overline{\epsilon} \sqrt{n+1} \, \chi_{n+1,j}) \,, 
\nonumber
\end{align}
define the transformation behavior of all 4d fields.

Given the mode expansions \eqref{matterKK} and \eqref{gaugeKK} it
is straightforward to obtain the full effective 4d action from the 6d
action \eqref{S6shifted}. The result reads
\begin{align}\label{S4shifted}
S_{4} = \int d^4 x \Big( &-\frac{1}{4}F_0^{\mu\nu} F_{0\mu\nu} 
- \partial^\mu \overline{\vf}_0\partial_\mu \vf_0 - \frac{1}{2}f^2\nonumber\\
+ \sum_{l,m}\Big(&-\frac{1}{4}F_{-l,-m}^{\mu\nu} F_{\mu\nu,l,m}
+ \frac{1}{2}\overline{M}_{-l,-m} M_{l,m} A^\mu_{-l,-m} A_{\mu,l,m} \nonumber\\
&- \partial^\mu \overline{\vf}_{l,m}\partial_\mu \vf_{l,m} 
- \frac{1}{4}\left|M_{-l,-m} \overline{\vf}_{-l,-m} + \overline{M}_{l,m} \vf_{l,m}\right|^2
\nonumber\\
&- \frac{i}{\sqrt{2}} A^\mu_{-l,-m}\partial_\mu
\left(M_{-l,-m} \overline{\vf}_{-l,-m} - \overline{M}_{l,m}
  \vf_{l,m}\right)\Big)\nonumber\\
+ \sum_{n,j}\Big(
&-i\psi_{n,j}\sigma^\mu \overline{D}_\mu \overline{\psi}_{n,j} - i\chi_{n,j}\sigma^\mu D_\mu \overline{\chi}_{n,j}\nonumber\\
& -\sqrt{2qf(n+1)}\chi_{n,j}\psi_{n+1,j} - \sqrt{2}q\vf_0 \chi_{n,j}\psi_{n,j} \nonumber\\
& -\sqrt{2qf(n+1)}\overline{\chi}_{n,j}\overline{\psi}_{n+1,j} 
- \sqrt{2}q\vfb_0 \overline{\chi}_{n,j}\overline{\psi}_{n,j}\Big) \nonumber\\
+\sum_{l,m; n,j; n',j'} &C^{l,m}_{n,j;n',j'} \Big(
-q \psi_{n',j'}\sigma^\mu A_{\mu,l,m} \overline{\psi}_{n,j} 
+ q \chi_{n,j}\sigma^\mu A_{\mu,l,m} \overline{\chi}_{n',j'} \nonumber\\
&\hspace{1.6cm}-\sqrt{2}q\vf_{l,m}\chi_{n,j}\psi_{n',j'} 
- \sqrt{2}q \overline{\vf}_{-l,-m}\overline{\chi}_{n',j'}\overline{\psi}_{n,j} \Big)\Big)\,.
\end{align}
Here the covariant derivatives $D_\mu$, $\overline{D}_\mu$ only involve the zero mode $A_{0\mu}$
of the gauge field, and the cubic couplings of the gauge and matter KK
modes are given by the overlap integrals
\begin{align}
C^{l,m}_{n,j;n',j'} = \int_{T_2} d^2 x \, \lambda_{l,m}
\overline{\xi}_{n,j}\xi_{n',j'}\,.
\end{align}
The action \eqref{S4shifted} describes the gauge modes $\vf_0$,
$\vf_{l,m}$, $A_{0\mu}$, $A_{\mu,l,m}$ and the fermion modes
$\psi_{n,j}$, $\chi_{n,j}$ as well as their interactions. In unitary gauge the
mixing between $A^\mu_{l,m}$ and one linear combination of $\vf_{l,m}$
and $\overline{\vf}_{l,m}$ is eliminated whereas the orthogonal
combination describes a tower of real, massive scalars.\footnote{In the supersymmetric
case this was discussed in \cite{Buchmuller:2016gib}.}

The mode functions of the charged matter fields are related by creation or
annihilaton operators, for instance (cf. \eqref{aplus},
\eqref{axiplus}),
\begin{align}
\overline{\xi}_{n-1,j} = -\frac{i}{\sqrt{n}} a_-\overline{\xi}_{n,j} 
= \frac{1}{\sqrt{2nqf}} (\pdb + qfz)\overline{\xi}_{n,j} \,.
\end{align}
Using expressions of this kind and performing partial integrations one
easily derives the following relations between the cubic couplings:
\begin{equation}\label{Crelations}
\begin{split}
\sqrt{2qf}\left(\sqrt{n} C^{l,m}_{n-1,j;n',j'} - \sqrt{n'+1}
  C^{l,m}_{n,j;n'+1,j'}\right) &= \overline{M}_{l,m} C^{l,m}_{n,j;n',j'}\,,\\
\sqrt{2qf}\left(-\sqrt{n+1} C^{l,m}_{n+1,j;n',j'} + \sqrt{n'}
  C^{l,m}_{n,j;n'-1,j'}\right) &= -M_{l,m} C^{l,m}_{n,j;n',j'}\,.
\end{split}
\end{equation}
With these relations it is straightforward to prove the invariance of
the action \eqref{S4shifted} under the transformations
\eqref{deltapsi}, \eqref{deltachi}, \eqref{deltaphi0},
\eqref{deltaphiprime} and \eqref{deltaAprime}. For instance, for the
Yukawa interactions one has
\begin{align}
\delta \sum_{l,m;n,j;n',j'} &C^{l,m}_{n,j;n',j'}
\vf_{l,m}\chi_{n,j}\psi_{n',j'} \nonumber\\
= \Big(&-\left(\bar{\epsilon}\overline{M}_{l,m} - \epsilon
  M_{l,m}\right) C^{l,m}_{n,j;n',j'} \nonumber\\
&+\sqrt{2qf}\left(-\epsilon \sqrt{n+1} C^{l,m}_{n+1,j;n',j'} 
+ \bar{\epsilon} \sqrt{n} C^{l,m}_{n-1,j;n',j'}\right) \nonumber\\ 
&-\sqrt{2qf}\left(\epsilon \sqrt{n'} C^{l,m}_{n,j;n'-1,j'} 
+\bar{\epsilon} \sqrt{n'+1} C^{l,m}_{n,j;n'+1,j'}\right)\Big)\ \vf_{l,m}\chi_{n,j}\psi_{n',j'}\nonumber\\
= \Big(&-\bar{\epsilon}\left(\overline{M}_{l,m} C^{l,m}_{n,j;n',j'}
+\sqrt{2qf}\left(\sqrt{n} C^{l,m}_{n-1,j;n',j'} - \sqrt{n'+1} C^{l,m}_{n,j;n'+1,j'}\right)\right)\nonumber\\
&+ \epsilon\left(M_{l,m} C^{l,m}_{n,j;n',j'} 
-\sqrt{2qf}\left(\sqrt{n+1} C^{l,m}_{n+1,j;n',j'} - \sqrt{n'}
  C^{l,m}_{n,j;n'-1,j'}\right)\right)\nonumber\\
&\quad \times \vf_{l,m}\chi_{n,j}\psi_{n',j'} \,.
\end{align}
Using Eqs.~\eqref{Crelations} one finds 
\begin{align}
\delta \sum_{l,m;n,j;n',j'} C^{l,m}_{n,j;n',j'}
\vf_{l,m}\chi_{n,j}\psi_{n',j'} = 0\,.
\end{align}
In the same way one shows the invariance of the other cubic
and bilinear terms in the action \eqref{S4shifted} involving gauge KK
modes. For the invariance of the remaining terms, which was already 
demonstrated above, the shift \eqref{deltaphi0} of the zero mode is
crucial, $\delta\vf_0 = \sqrt{2}\bar{\epsilon} f$.

It is instructive to contrast this result with the generation of a mass for the WL scalar in gauge-Higgs unification without flux. The inverse size of the torus plays the role of a cutoff, and a discrete symmetry, a remnant of the gauge symmetry in the compact dimensions, keeps the mass finite. Explicit expressions for the effective potential of the WL scalar have been obtained in Refs. \cite{Antoniadis:2001cv, Ghilencea:2005vm}. For a square torus of size $L = 2 \pi R$ one obtains \cite{Dierigl:2017sqi}
\begin{align}
m_{\varphi_0}^2 \approx 0.19 \times  \frac{\alpha}{\pi} \frac{1}{R^2} \,.
\end{align}
As expected, the result is the product of a loop factor and the cutoff. Turning on flux changes the situation drastically. The full 4d action including the couplings of all KK modes is invariant with respect to a symmetry under which the scalar $\vf_0$ shifts by a constant. The origin of this symmetry is the shift of $\vf$ under translations on the torus, which compensates for the spontaneous breaking of translation invariance by the flux potential. As a result, the mass of the WL scalar vanishes.

\section{Summary and Outlook}

Motivated by the hierarchy problem of the electroweak theory we have
studied the effect of magnetic flux on quantum corrections to a scalar
mass term in a model of gauge-Higgs unification. We considered the
simplest possible example, a 6d Weyl fermion with Abelian gauge
interaction, compactified on a torus. 

We first analyzed the symmetries of the 6d theory on a torus. In the
presence of the background gauge field translational invariance is
realized non-linearly, and the Higgs field transforms with a
shift. Moreover, the theory has a well known local 6d symmetry 
under which the Higgs field also transforms with a shift, and which
changes the boundary conditions of the charged fields. Using the
familiar harmonic oscillator algebra a complete orthonormal set
of mode functions was constructed for the two 4d Weyl fermions
with opposite charge, which are contained in the 6d Weyl fermion.
The Higgs field and the vector field have the standard Kaluza-Klein
mode expansion. The effective 4d action contains as zero modes
a multiplicity of Weyl fermions, determined by the magnetic flux, 
and a complex scalar with chiral couplings to pairs of
different fermions.

Quantum corrections were studied in three steps. Keeping just the
lowest lying massive Dirac fermion, a quadratic divergence for the
scalar mass term is generated at one loop, as expected. However,
once the full tower of massive states is included, the total
correction to the scalar mass term vanishes, confirming our earlier
result. The origin of this cancellation of quantum corrections is a
symmetry of the 4d effective action. Starting from the two symmetries
of the 6d theory, translations and gauge invariance, we identified a
transformation law of the 4d fields which leaves the 4d action
invariant. Under this transformation the complex scalar transforms
with a shift, which prevents the generation of a mass term.

In a third step we generalized this result to the complete 4d theory,
including the Kaluza-Klein excitations of scalar and vector fields.
Using properties of the mode functions we obtained relations among
the cubic couplings, which allowed us to demonstrate that the full 4d theory
has an exact symmetry under which the complex Higgs field transforms
with a shift. The origin of this shift symmetry are the translation
symmetries of the torus. Assuming a renormalization scheme, which
preserves this symmetry, we conclude that no scalar mass term will
be generated at any loop order. This is the main result of the present
paper.

What is the relevance of this result for the hierarchy problem of the electroweak theory? The effective low energy action of our model was given in Section~\ref{sec:torusflux}. The action has a shift symmetry, and a vacuum expectation value of the scalar field does not generate mass terms for chiral fermions, which is the key feature of the Standard Model. In order to obtain more realistic low energy models one has to consider non-Abelian gauge theories in higher dimensions. It is then possible to obtain flux compactifications with gauge-Higgs unification where the Higgs field couples to chiral fermions and all fields have Landau-level excitations. It remains to be seen whether also in these theories, where the Higgs field is charged, an approximate shift symmetry can be realized once the contribution of Kaluza-Klein excitations is taken into account. These questions are currently under investigation.

\section*{Acknowledgments}
We thank Constantin Bachas, Arthur Hebecker, Miguel Montero, and Augusto Sagnotti for
valuable discussions. This work was supported by the German Science
Foundation (DFG) within the Collaborative Research Center (SFB) 676
``Particles, Strings and the Early Universe''. M.D.'s work is part of
the D-ITP consortium, a program of the Netherlands Organisation for
Scientific Research (NWO) that is funded by the Dutch Ministry of
Education, Culture and Science (OCW). E.D. was supported in part by the ``Agence Nationale de la Recherche" (ANR).

\appendix

\section{Quantum mechanics on the magnetized torus}
\label{app:qm}

It is possible to understand the properties under translations on the magnetized torus by using elementary quantum mechanics arguments (see e.g.\ \cite{Abouelsaood:1986gd, Tong:2016kpv}). In the presence of a
magnetic field, $F_{56} = f$, the gauge potential in the two directions of the torus, in the symmetric gauge we used in the previous sections, is given by
\begin{equation}
A_5 = - \frac{1}{2} f x_6 + a_5 \,, \quad A_6 = \frac{1}{2} f x_5 + a_6 \,, \label{qm1} 
\end{equation}
where $a_{5,6}$ are constants (i.e.\ Wilson lines). The non-trivial transformations of the gauge potentials under translations along the two cycles of the torus are
\begin{equation}
A_5 (x_5, x_6 + \epsilon_6) =  A_5 (x_5,x_6) -  \frac{1}{2} f \epsilon_6 \,, \quad  A_6 (x_5 + \epsilon_5, x_6) =  A_6 (x_5,x_6) +  \frac{1}{2} f \epsilon_5 \ , \label{qm2} 
\end{equation} 
which can also be described as gauge transformations with parameters,
\begin{equation}
\Lambda_5 = \Lambda_5^{(0)} +  \frac{1}{2} f \epsilon_5 \, x_6  \,, \quad \Lambda_6 = \Lambda_6^{(0)} -  \frac{1}{2} f \epsilon_6 \, x_5 \,,   \label{qm3} 
\end{equation}
where $\Lambda_{5,6}^{(0)}$ describe constant gauge transformations which allow us to eliminate trivial (by constant factors) transformations of the wave functions.
The Hamiltonian $H$ of a particle of charge $q$ on the magnetized torus is given by
\begin{equation}
H = \frac{1}{2} \left(P_5 - \frac{q f}{2} x_6 + q a_5 \right)^2 +
\frac{1}{2} \left(P_6 +  \frac{q f}{2} x_5 + q a_6\right)^2  
\ , \label{qm4} 
\end{equation}
where $P_5,P_6$ are the momenta, acting on the wave functions in the
standard way $P_{5,6} = - i \partial_{5,6}$. The operators related to
translations by $\epsilon_{5,6}$ on the torus in a flux background have to commute
with the Hamiltonian \eqref{qm4}. They are explicitly given by
\begin{equation}
\Pi_5 = e^{i \epsilon_5 (P_5 + \frac{qf}{2} x_6) } \,, \quad  \Pi_6 = e^{i \epsilon_6 (P_6 - \frac{qf}{2} x_5) } \,. \label{qm5} 
\end{equation}
Since they commute with $H$ one can choose wave functions which are
eigenvectors of the 
translations operators. 
From \eqref{qm5} or \eqref{qm3} one can derive the
transformation behavior of wave functions of charged fields under lattice translations
\begin{eqnarray}
&& \Psi (x_5 + L, x_6) \ = \ e^{- \ttfrac{i}{2} q f L x_6  + i q L a_5} \  \Psi (x_5, x_6) \ , \nonumber \\
&& \Psi (x_5 , x_6 + L) \ = \ e^{\ttfrac{i}{2} q f  L x_5 + i q L a_6 } \ \Psi (x_5, x_6)  \ . \label{qm6} 
\end{eqnarray}
Notice that by taking closed loops around the two cycles,
$\epsilon_{5,6} = L$, and imposing single-valuedness of the  wave function, one can derive the quantization of the magnetic flux 
\cite{Abouelsaood:1986gd,Bachas:1995ik}. By defining complex
translations with
\begin{equation}
\epsilon = \frac{1}{2} (\epsilon_5 + i \epsilon_6)  \,, \label{qm7} 
\end{equation}
one obtains a complex translation operator on the torus implementing
$z \rightarrow z + \epsilon$.  Using the Campbell-Hausdorff formula one finds
\begin{equation}
\begin{split}
\Pi_{\epsilon} =  \Pi_{6} \Pi_{5} &= \exp \Big(\epsilon \partial_z + {\bar \epsilon} {\bar \partial}_z  - q f  (\epsilon {\bar z} - {\bar \epsilon} z) -  \frac{q f}{2} (\epsilon^2- {\bar \epsilon}^2) \Big) \\
&= \exp \Big(- i \sqrt{ 2q f } \ ( \epsilon a^\dagger_-  + {\bar \epsilon} a_-) -  \frac{q f}{2} (\epsilon^2- {\bar \epsilon}^2) \Big)\,, \label{qm8}
\end{split}
\end{equation}
with $a_-$ and $a_-^\dagger$ defined in Eqs.~\eqref{aminus}.
This symmetry of the Hamiltonian is a standard one since, as we will
discuss below, $a_-$ and $a_-^\dagger$ are acting in the degenerate Fock space of Landau levels of a given mass,  creating the degeneracy
described by the quantum number $j$. 

On the other hand, the Schr\"odinger equation of a charged particle in the magnetic field has another, less obvious symmetry. Let us search for a symmetry of the Schr\"odinger equation,
\begin{equation}
H \Psi = E \Psi \,, \quad  H'  \Psi' = E \Psi'    \,, \label{qm9}
\end{equation}
which mixes states of different mass.  It is possible to realize this,
while keeping the energy eigenvalue $E$ invariant, if one also performs changes of the field $\varphi$. 
First of all, the Hamiltonian \eqref{qm4} can be written as 
\begin{equation}
H = qf \left(a_+^\dagger a_+ + \frac{1}{2}\right) + i q \sqrt{qf} (\varphi a_+^{\dagger} - {\overline\varphi} a_+) + q^2 |\varphi|^2   \ , \label{qm10}
\end{equation}
with $a_+$ and $a_+^\dagger$ defined in Eqs.~\eqref{aplus}.
A symmetry of the Schr\"odinger equation is of the form
\begin{equation}
\Psi' = U \Psi \,, \quad H' = U H U^{-1}  \ , \label{qm11}
\end{equation}
with a unitary operator $U$. For an infinitesimal transformation $U = e^{i T} \simeq 1 + i T$, with $T$ a hermitian generator, one finds
\begin{equation}
\delta \Psi = i T  \Psi  \,, \quad \delta H = i [T, H]  \,, \label{qm12}
\end{equation}
where in our case
\begin{equation}
\delta H = i q \sqrt{qf} (\delta \varphi a_+^{\dagger} -  {\delta \overline{\varphi}} a_+) + q^2 ( \varphi \delta {\overline \varphi} +   {\overline \varphi}  \delta \varphi )  \ . \label{qm13}
\end{equation}
It is then straightforward to verify that \eqref{qm12} is satisfied, with
\begin{equation}
\begin{split}
T &= - \sqrt{2 qf } (\epsilon  a_+ +  {\bar \epsilon} a_+^\dagger) \,, \\
\delta \varphi &=  \sqrt{2} \bar \epsilon f  \,. \label{qm14}
\end{split}
\end{equation}
Finally, the hidden and non-linearly realized symmetry of the Schr\"odinger equation acts as
\begin{equation}
\begin{split}
\delta \Psi &= - i \sqrt{2 qf} (\epsilon  a_+ +  {\bar \epsilon} a_+^\dagger ) \Psi \,, \\
\delta \varphi &= \sqrt{2} \bar \epsilon f  \,. \label{qm15}
\end{split}
\end{equation}
This is the quantum mechanical analog of the higher-dimensional
symmetry found in field theory, mixing all Landau mass levels, under
which the scalar $\varphi$ transforms as a Goldstone boson, see Eqs.~\eqref{trafopsi}, \eqref{trafochi} and \eqref{deltaphi0}.  In the quantum mechanical case, in which the gauge field is external (not quantized), this would just imply that the gauge potential
is unphysical and can be set to zero. 

Finally, we would like to briefly discuss the properties of
wave functions in the magnetic field in the symmetric gauge that we are
using (see, for example \cite{Tong:2016kpv}). One can introduce an
angular momentum operator\footnote{The negative sign in its definition is a
 matter of convention.} 
\begin{equation}
- J = x_5 P_6 - x_6 P_5 = z \partial_z - {\bar z} \partial_{\bar z} =  a_+^{\dagger} a_+ - a_-^{\dagger} a_- \ . \label{qm16}
\end{equation} 
 Notice that $[H_0, J]= 0$  and $[J, \Pi_{\epsilon}] \not=0$, where
 $H_0 = H (\varphi=0)$. Hence, one can choose wave functions with definite angular momentum or eigenvectors of the translation operator, but in general not both simultaneously. The usual choice are states of definite angular momentum $j$. In this case, $a_-$, $a_-^\dagger$ generate the Fock space of states $| n,j\rangle $, $j = 0,\dots,N-1$ for a given 
 oscillator quantum number $n$, where $N$ is the magnetic field flux.  Indeed, notice in particular that, in non-compact space, 
 \begin{equation}
 | 0 , j \rangle  =  \frac{(a_-^\dagger)^j}{\sqrt{j !}}  | 0 , 0  \rangle \sim {\bar z}^j e^{-q f |z|^2} \ , \label{qm17}   
  \end{equation}
 with corresponding wave functions in the compact space constructed by adding images, as usual. 
   The Fock space of wave functions can be explicitly constructed according to
 \begin{equation}
 | n , j \rangle = \frac{(a_+^\dagger)^n}{\sqrt{n !}}  \frac{(a_-^\dagger)^j}{\sqrt{j !}}  | 0 , 0 \rangle  \ . \label{qm17}
 \end{equation}
 
 }
 
% #################################
% #           Bibliography        #
% #################################
\providecommand{\href}[2]{#2}\begingroup\raggedright\endgroup

\end{document}